\documentclass[pra,12pt]{revtex4}

\newcommand{\Vector}{\mathbf}

\usepackage{braket}		
\usepackage{amsmath}    
\usepackage{graphicx}   
\usepackage{verbatim}   
\usepackage{color}      
\usepackage{subfigure}  
\usepackage{hyperref}   

\begin{document}

\title{Closed Form of the Biphton K-Vector Spectrum for Arbitrary Spatio-Temporal Pump Modes}
\author{Jeffrey Perkins}
\date{\today}
\begin{abstract}
A closed form solution is derived for the biphoton k-vector spectrum for an arbitrary pump spatial mode.  The resulting mode coefficients for the pump input that maximize the probability of biphoton detection in the far field are found.  It is thus possible to include the effect of arbitrary crystal poling strucures, and optimize the resulting biphoton flux.
\end{abstract}

\maketitle

\section{Introduction}  

The effect of higher order or multiple pump spatial modes on the shape of the biphoton wavepacket resulting from spontaneous parameteric downconversion (SPDC) is of interest for several reasons, including the determinsitic (heralded) generation of arbitrary single photon states \cite{jeffrey2004towards}, entanglement of large numbers of photons \cite{eisenberg2004generation}, super-resolving phase measurements \cite{mitchell2003super}, and the entanglement of differing angular momentum states of the electromagnetic field \cite{walborn2004entanglement}.  

Previous theoretical \cite{lantz2004spatial, ling2008absolute} and experimental \cite{castelletto2004measurement, molina2005control} investigations into the spatial distribution of the quantum noise, in SPDC were rendered incomplete due to the inability to determine the biphoton spectra of higher order TEM modes.  From a theoretical viewpoint, an experimentally testable form for the biphoton spectra of an arbitrary spatially shaped pump pulse has not been derived.  Experimentally, the complete impact of differing experimental parameters on the corresponding parameters of the biphoton wavepacket is desired.  We propose to solve these problems by choosing a different basis with a defined Fourier Transform and solving for a complete, closed form solution with Siegman's Elegant Gauss Hermite Modes, while only assuming that the width of the crystal when compared to the Guassian spot size is large such that the transverse Fourier Transform may be analyzed.  

In short, the ``Elegant Gauss Hermite Modes,'' also known as the ``Complex Gauss Hermite Modes'' (which we will use interchangeably throughout) have a defined, derivable Fourier Transform.  This is due to the fact that they may be written as total derivatives acting on the paraxial diffraction kernel.  We will use this alternative mode expansion to convert the interaction Hamiltonian of a single nonlinear crystal from an integral over the crystal volume to an integral over the momenta of the signal and idler photons, thus deriving the shape of the biphoton spectral density for an arbitrary superposition of Complex Gauss Hermite Modes.  The form of the biphoton spectral density for arbitrary spatial pump distributions has not been derived to the author's knowledge.  It is thus impossible to tailor a pump beam to optimize the biphoton distribution along anything but the optic axis of the pump laser without this solution.  Although we will not do so here, in the far field, the biphoton spatial frequencies may be interpreted as the angle that the signal or idler photon makes with the pump beam optic axis; we may thus convert the integral over a corresponding detector area into an equivalent integral over the small region of spatial frequencies that are sampled by a detector with limited area.  

We will start by familiarizing the reader with the notation to follow.  The notation is included in a helpful table.  

To simplify the notably variable laden integrals to follow, we will use $\xi \equiv 1 + i\frac{z}{z_r}$, following Boyd \cite{boyd2003nonlinear}, and $\rho^{2} \equiv x^2 + y^2$ throughout.  $u_0(t)$ is a normalization constant from Seigmann that will later represent the time variation of the amplitude (it will remain classical, as the pump field is treated classically; technically, it is ``quantized'' and we assume a coherent single mode, albeit Elegant Gauss-Hermite, and replace the raising and lowering operators with their eigenvalues) \cite{siegman1973hermite}.  Sums are explicitly expressed.  $\Vector{r}_{\bot}$ is the component of r that is perpendicular to the optical axis (by convention, we define positive ``z'' to be the optical axis of the pump laser).  Also, $\Vector{\nu}_+ \equiv \Vector{\nu}_s + \Vector{\nu}_i$, or the sum of the signal and idler fields that will be defined later.  Any subscript ``+'' is to represent the sum of the signal and idler variables of that particular variables.  We will also use $\epsilon_{s}$ to represent the two dimensional polarization vector of the electromagnetic field.  We point out that choosing $\xi \equiv 1 + i\frac{z}{z_r}$ rather than $\xi \equiv 1 - i\frac{z}{z_r}$ is equivalent to choosing the form $\Vector{k} \cdot \Vector{r} - \omega t$ for the phase of the electromagnetic field, rather than its negative. 
\\
\\

$\begin{array}{|c|c|}\hline 
w_{0} & Laser \ \ Spot \ \ Size \\\hline 
\xi &  1 + i\frac{z}{z_r} \\\hline 
\rho^{2} & x^2 + y^2 \\\hline
\Vector{r}_{\bot} & x\hat{x} + y\hat{y}\\\hline 
z_{r} & \pi w_{0}^{2}/{\lambda}\\\hline   
u_0(t) & Time\ \ Varying\ \ Amplitude \\\hline
z & Pump\ \ Optic\ \ Axis \\\hline
\Vector{\nu}_{+} & \Vector{\nu}_{s} + \Vector{\nu}_{i} \\\hline
\epsilon_{s} & Polarizaton\ \ Vector \\\hline 
 \end{array}$
\\
\\
An elegant Gauss-Hermite mode is  defined as \cite{siegman1973hermite}:
\begin{align}
u_{nm} &= u_0 \frac{(-w_0)}{\xi}^{n+m} \frac{d^n d^m}{dx dy}e^{-\frac{\pi \rho^2}{\lambda z_r \xi}}
\end{align}

The derivation from Siegman's original form in \cite{siegman1973hermite} to the form stated above is contained in the Appendix.

An elegant Gauss-Hermite mode of order n,m is biorthogonal with the function $\psi_{nm}$:\cite{siegman1973hermite}:
\begin{align}
\iint^{\infty}_{-\infty} dxdy\ \ u_{nm} \psi_{nm} &=  \delta_{nm}f_{nm}(z)\\
\psi_{nm} &= H_n\left({\sqrt{\frac{\pi}{\lambda z_r \xi^*}}x}\right)H_m\left({\sqrt{\frac{\pi}{\lambda z_r \xi^*}}y}\right) 
\end{align}

Although the Hermite polynomials here are complex, and in fact the solution to a non-Hermitian differential equation, we may reasonably expect that a physicaly generated laser beam (such as might be expected for the pump beam) will show a distribution that asymptotically converges to zero as the transverse area tends to infinity.  This is equivalent to assuming that the total energy of the pump beam is finite.  If this convergence is Gaussian, as is common, and we set the origin of the Complex Gauss Hermite Modes to the same origin as the pump beam, we can see that we may assume that the resulting polynomial expression in front of the Gaussian phase factor will be insufficient to disrupt the aforementioned convergence.  In short, we can assume that, for a physical laser beam generated in a cavity (as is common for high power Q-switched lasers that are commonly used as the pump lasers in these experiments), the resulting spatial profile will be a superposition of ``regular'' Gauss Hermite modes, and the multiplication of two real Hermite polynomials is not sufficiently divergent to overcome the gaussian damping factor.  This is possible because we choose the origin of the Elegant Gauss Hermite Modes such that the overlap integral is performed at that origin, the only position where the modes must be real.  The factor $\xi$ will be 1 at that origin.

The $f_{nm}(z)$ is a constant when the origin of the elegant modes corresponds to the origin of the physical electric field \cite{siegman1973hermite}.
\begin{align}
u_{nm} &= u_0(t) \frac{(-w_0)^{n}(-w_{0})^{m}}{\xi}\frac{d^n d^m}{dx dy}e^{-\frac{\pi \rho^2}{\lambda z_r \xi}}
\end{align}

We may write the electric field in its entirety as, with l as the polarization index:
\begin{align}
E_{p} (\Vector{r}, t) = e^{2\pi i\nu_{p}z}\sum_{n, m, l} u_{nml}(t) c_{nm}\epsilon_{l}  + C.C.
\end{align}
  
This form of the electric field satisfies the paraxial wave equation.  This can be seen by direct application of the paraxial wave equation.  The paraxial wave equation is:
\begin{align}
\psi(\Vector{r})  = u(\Vector{r}) e^{ikz}
\nabla^{2}_{\bot} u(\Vector{r}) + 2ik \frac{\partial u(\Vector{r})}{\partial z} = 0
\end{align}

We now find the Fourier Transform of the elegant Gauss-Hermite modes; we list all relevant Fourier Transforms that are required to analyze the modes in this manner:
\begin{align}
FT (f(\Vector{r})) = F(\nu_x, \nu_y, \nu_z )&=\iiint_{-\infty}^{\infty} d^3x\ \ f(\Vector{r}) e^{-2\pi i \Vector{\nu} \cdot \Vector{r}}\,\\
 FT_{\bot} (f(\Vector{r})) = F(\nu_x, \nu_y, z )&=\iint_{-\infty}^{\infty}dxdy\ \ f(\Vector{r}) e^{-2\pi i \Vector{\nu_{\bot}} \cdot \Vector{r}}\, \\
 \label{eq:FTderiv}
FT_{\bot} \left(\frac{d^{n}d^{m} f(x/a)f(y/b)}{dx^n dy^m}\right) &= ab (2\pi i\nu_x)^n (2\pi i\nu_y)^m F(\nu_x, \nu_y)\\
FT_{\bot} \left(e^{-\pi\left((x/a)^2+(y/b)^2\right)} \right) &= ab e^{-\pi\left((a\nu_x)^{2} + (b\nu_y)^{2}\right)}\\
\label{FT_EGH}
\end{align}

A brief note of importance here: first, \eqref{eq:FTderiv} does \emph{not} have the scaling factor that is normally present in the spatial frequencies; this is expected for two reasons: first, an analysis of the units associated with the Fourier Transform demonstrates that if the scaling factors were present, the result would be dimensionally incorrect.  Second, the scaling factor cancels in the following manner, which we will demonstrate in one dimension:
\begin{align}
x' &\equiv x / a\\
dx' &= dx / a\\
\frac{d}{dx'} &= a^{-1} \frac{d}{dx}\\
FT\left(\frac{d^{n}}{dx^{n}} f(x/a)\right) &= \int dx \frac{d^{n}}{dx^{n}} f(x/a) e^{i 2\pi\nu_{x}x}\\
FT\left(\frac{d^{n}}{dx^{n}} f(x/a)\right) &= a^{-n}\int (a dx') \frac{d^{n}}{dx'^{n}} f(x') e^{i 2\pi (a \nu_{x}) x'}\\
FT\left(\frac{d^{n}}{dx^{n}} f(x/a)\right)  &= \frac{i2\pi \nu_{x} a}{a^{n}} a F(a\nu_{x})\\
FT\left(\frac{d^{n}}{dx^{n}} f(x/a)\right)  &= a(i2\pi \nu_{x})^{n} F(a\nu_{x})
\end{align}

This fact will allow us to completely derive the form of the interaction Hamiltonian in three dimensions, resulting in a derivable phasematching function (else the Fourier Transform in z would not have a closed form).

Relating these to the Fourier Transform of the Complex Gauss Hermite modes, we see that we have a function of the form $\frac{d^{n}d^{m} e^{-\pi\left((x/a)^2+(y/b)^2\right)}}{dx^{n}dy^{n}}$, where a and b are, for a circular Complex Gauss Hermite mode, equal.  They are both (for a circular case) equal to $\sqrt{\lambda z_{r} \xi}$, and the multiplication of the two (ab) will give $\lambda z_{r} \xi$, which will, in combination with the prefactor $\frac{(-w_{0})^{n + m}}{\xi}$, cancel the $\xi$ on the denominator.  This will allow us to evaluate a simple longitudinal integral, which is merely a plane wave.  Noting that $\lambda z_{r}$ is $\pi w_{0}^{2}$, or the transverse area of the beam, and collecting like terms, the Fourier transform is: 
\begin{align}
FT (u_{nm}) &= u_0(t)A_{\bot}\left(-2\pi i w_0 \nu_x \right)^{n} \left(-2\pi i w_0 \nu_y \right)^{m} e^{-\pi \lambda z_r \xi(\nu_x^2 + \nu_y^2)} 
\end{align}

$A_{\bot} \equiv \pi  w^2_{0}$ is the transverse area of the beam (actually, this is the $e^{-2}$ area of the beam, or approximately 95\% of it).  These results are expressed in a form sufficient for generalization to elliptical Elegant Gauss-Hermite modes (simply, the spot sizes in two dimensions need not be the same, and are merely scaling factors for the spatial frequencies in the appropriate dimension.).

We will use \eqref{FT_EGH} in the derivation of the interaction Hamiltonian.

\section{Derivation of the Elegant Gauss Hermite Interaction Hamiltonian}
The derivation of the interaction Hamiltonian will follow quite simply from the interaction Hamiltonian for three wave mixing \cite{shih2003entangled}:
\begin{align}
H_{i}(\Vector{r}, t) = \epsilon_0 \int_V{ d^3r\chi(\Vector{r})^{s_{p}s_{i}s_{s}} {E^{s_{p}}_p(\Vector{r}, t}){E^{s_{s}}_s(\Vector{r}, t)}{E^{s_{i}}_i}(\Vector{r}, t)}
\end{align}

We will treat the time, and thus spectral, dependence of the field in a following section.  For now, it is understood that we will only determine the spatial frequency dependence of the interaction.  We will also treat the differing indices of refraction, inherently included in $\Vector{\nu} \equiv \Vector{\nu}_{fs} / n$, where ``fs'' stands for ``free space'', for the calculation of the interaction Hamiltonian.

For simplicity, we will treat the nonlinear crystal as cut for type-II (for example, o-eo) phase matching with a large, essentially infinite, surface area that is normal to the pump beam optical axis, but a finite length along the optic axis; the assumption of the infinite transverse area is our only assumption.  Such an assumption is common in nonlinear optics \cite{boyd2003nonlinear}.  As such, we may assume that the raising and lowering operators of the signal and idler modes always commute.  We may also drop the polarization vector, as the polarization of the signal and idler modes is now orthogonal, and set the overall phase of the pump mode to zero.  

We will also assume a coherent state for the pump field such that it may be treated classically.  We also assume that the nonlinear coefficient does not vary over the crystal length for simplicity.

We will write the fields as:
\begin{align}
\Vector{E_{s}}(\Vector{r}, t) &= \sqrt{\frac{2h}{\epsilon_{0}}}\sum_{s}\int d^{3}\sqrt{f_{s}}\nu_{s} \left[ia_{s}(\Vector{\nu_{s}})\Vector{\epsilon}_{s}e^{i2\pi(\Vector{\nu_{s}} \cdot \Vector{r} - f_{s} t)} + H.C.\right]\\
\Vector{E_{i}}(\Vector{r}, t) &= \sqrt{\frac{2h}{\epsilon_{0}}}\sum_{s}\int d^{3}\sqrt{f_{i}}\nu_{i} \left[ia_{i}(\Vector{\nu_{i}})\Vector{\epsilon}_{s}e^{i2\pi(\Vector{\nu_{i}} \cdot \Vector{r} - f_{i }t)} + H.C.\right]\\
\end{align}

Where $\nu$ is the spatial frequency, m is the index of polarization, and f is the frequency of the photon; as such, $hf$ is the energy of the photon, where h is Planck's constant.  The fields are quantized in free space, and the nonlinear crystal is treated as a perturbative region of interaction.  We note that we chose to operate in $\Vector{\nu}$-space rather than $\Vector{k}$-space as normalization constants of the Fourier Transform are unity in $\Vector{\nu}$-space, and this leads to a more elegant form of the interaction Hamiltonian.

For clarity, we now drop the polarization index; we assume that the signal and idler fields are perpendicularly polarized (Type-II SPDC) and thus commute; it is possible to solve the problem with frames of vector spaces for identically polarized signal and idler fields with a frequency overlap, which would constitute an overcomplete basis.  However, for the purposes of clarity, we demonstrate the (new) technique here without such complications.  Explicitly, $[a_i, a_s^{\dagger}] \equiv 0$.

The quantum mechanical interaction Hamiltonian of interest for SPDC is proportional to $a^{\dagger}_s a^{\dagger}_i$ and reduces to a Fourier Transform \cite{shih2003entangled}:
\begin{equation}
H_{SPDC} = -2h\chi \sum_{n, m} c_{nm} \int_V\ d^3r d^3\nu_s d^3\nu_i u_{nm}\sqrt{f_s(\nu_s) f_i(\nu_i)}e^{i\frac{2\pi}{\lambda}z} a^{\dagger}(\nu_s) a^{\dagger}(\nu_i) e^{-i2 \pi (\nu_s + \nu_i) \cdot r)}
\end{equation}
We now take the Fourier Transform over the transverse area to find:
\begin{equation}
H_{SPDC} = 2h \chi A u_0(t) \sum_{n, m} c_{nm}
\int_V\ dz d^3\nu_s d^3\nu_i \left\lbrace
\begin{aligned} 
&a^{\dagger}(\nu_s) a^{\dagger}(\nu_i) \sqrt{f_s(\nu_s) f_i(\nu_i)} e^{-i2 \pi (\nu_{+z} z)} e^{i\frac{2\pi}{\lambda}z} \\
&\times e^{-i\pi\lambda z \left(\nu_{+x}^2 + \nu_{+y}^2 \right)} e^{-\pi\lambda z_r\left(\nu_{+x}^2 + \nu_{+y}^2 \right)}\\
&\times (2\pi iw_0 \nu_{+x})^n(2\pi iw_0\nu_{+y})^m 
\end{aligned}
\right\rbrace
\end{equation}

Where we have made use of \eqref{FT_EGH}.  The overall minus sign results from convention.

Note here that we were careful to ensure that the complex phase of the pump mode, $~e^{+ik_{p}z}$ and the gaussian phase factor, was positive, in keeping with our convention of defining the phase of the lower operator of the field in the same way.  This was a direct result of treating the state of the pump field semi-classically; the complete Hamiltonian would include a term proportional to $a_{p}a^{\dagger}_{s}a^{\dagger}_{i}$, and the coherent state input would merely replace $a_{p}$ with the appropriately scaled amplitude of the pump.  In keeping with our intuition, integration over all space would return a delta function where the sum of the signal and idler k-vectors along the optics axis must be identical to the initial pump mode optical axis k-vector.

If we label the momentum mismatch $(\frac{1}{\lambda} - \lambda \nu_{+\bot}^2 -\nu_{+z}) \equiv \Delta \nu$, we find evaluation over the longitudinal, and thus collinear with the optical axis, dimension, $z$, gives the expected phasematching function \cite{boyd2003nonlinear} $\Phi ={e^{i2\pi \Delta \nu \Delta z} \frac{\sin( \pi \Delta \nu L)}{ \pi L \Delta \nu}}$, multiplied by the length of the interaction L,  where $\Delta z$ is positive for a Gaussian mode with a focal mode that corresponds to the back of the crystal.  We will call V the volume of the interaction, defined for the moment as AL, where A is the transverse area of the beam, $\pi w^{2}_0$, and L is nominally the crystal length, although properly it is the length over which the interaction is coherent.
\begin{align}
H_{SPDC} &=  -2h \chi V u_0(t) \sum_{n, m} c_{nm} \int
d^3\nu_s d^3\nu_i\left\lbrace
\begin{aligned} 
&a^{\dagger}(\nu_s) a^{\dagger}(\nu_i) \sqrt{f_s(\nu_s) f_i(\nu_i)} (2\pi iw_0 \nu_{+x})^n\\
& \times (2\pi iw_0\nu_{+y})^m \Phi(\Delta \nu) e^{-\pi\lambda z_r\left(\nu_{+x}^2 + \nu_{+y}^2 \right)} 
\end{aligned} \right\rbrace
\end{align}

\section{Time Evolution Operator and the Biphoton State}

In this section, we derive the time evolution operator to first order in the nonlinear coefficient $\chi$.  It is defined as \cite{fetter2003quantum}:
\begin{equation}
U(t, t_0) \equiv \sum^{\infty}_{n=0} \left( \frac{-i}{\hbar}\right)^n \frac{1}{n!} \int^{t}_{t_0} dt_{1} \int^{t}_{t_0} dt_{2} \cdots \int^{t}_{t_0} dt_{n} \ \ T[ H_i(t_1) H_i(t_1) \cdots H_i(t_n) ]
\end{equation}

We insert the time dependence of the pump field into the amplitude $u_0(t)$; the time dependence of the raising operators is well known \cite{fetter2003quantum}.   With this substitution, we find $H_{SPDC}(t) \propto  u_0(t) e^{i 2\pi t f_{+}}$ where $f_{+} \equiv f_s(\nu_s) + f_i(\nu_i)$.

We treat the nonlinear coefficient $\chi$ as small and expand the propagator to first order to find, noting the Fourier transform of $u_0(t)$ is denoted as $\widetilde{u}(f)$:
\begin{align}
U(t, t0) &\approx 1 - \frac{i}{\hbar} \int^t_{t_0} dt' H_{I}(t')\\
U(\infty, -\infty) &\approx  
 2h \chi V u_0 \sum_{n, m} c_{nm}\int
d^3\nu_s d^3  \nu_i \widetilde{u}(f_+)\left\lbrace
\begin{aligned} 
&a^{\dagger}(\nu_s) a^{\dagger}(\nu_i) \sqrt{f_s(\nu_s) f_i(\nu_i)} (2\pi iw_0 \nu_{+x})^n\\
& \times (2\pi iw_0\nu_{+y})^m \Phi(\Delta \nu) e^{-\pi\lambda z_r\left(\nu_{+x}^2 + \nu_{+y}^2 \right)} 
\end{aligned} \right\rbrace
\end{align}

We may thus derive for the biphoton state vector, noting that the initial state is the vacuum state, and assuming the interaction is "turned on" for a time approaching infinity, the following state:
\begin{align}
\ket{\psi_I (t = \infty)} &= U(\infty, -\infty) \ket{0}\\
\label{Answer}
\ket{\psi_I (t = \infty)} &=  2h \chi V u_0 \sum_{n, m} c_{nm}\int
d^3\nu_s d^3  \nu_i \widetilde{u}(f_+)\left\lbrace
\begin{aligned} 
&(2\pi iw_0 \nu_{+x})^n (2\pi iw_0\nu_{+y})^m \Phi(\Delta \nu) \\
&\times  e^{-\pi\lambda z_r\left(\nu_{+x}^2 + \nu_{+y}^2 \right)} \sqrt{f_s(\nu_s) f_i(\nu_i)}  
\end{aligned} \right\rbrace
\ket{\nu_i}\ket{\nu_s}
\end{align}

With rare mathematical exceptions that might be separable, the multimode state is thus obviously entangled.  We now proceed to derive a similar Hamiltonian, but include the sum over multiple polarizations and multiple crystals of the same length pumped by the same beam.

\section{Generalized Hamiltonian for Spontaneous Downconversion}

If we wish to include polarization, we simply insert the polarization index into both $\chi$ and the polarization of the fields.  We then find:

\begin{align}
H_{SPDC} &=  -2h\sum_{n, m, o, q, r} c_{nm} \chi^{oqr} \epsilon_{po}\epsilon^{*}_{sq}\epsilon^{*}_{ir} V u_0  \int
d^3\nu_s d^3\nu_i\left\lbrace
\begin{aligned} 
&a^{\dagger}_{sq}(\nu_s) a^{\dagger}_{ir}(\nu_i) \sqrt{f_s(\nu_s) f_i(\nu_i)} (2\pi iw_0 \nu_{+x})^n\\
& \times (2\pi iw_0\nu_{+y})^m \Phi(\Delta \nu) e^{-\pi\lambda z_r\left(\nu_{+x}^2 + \nu_{+y}^2 \right)} 
\end{aligned} \right\rbrace
\end{align}

Where $\epsilon_{po}$, $\epsilon^{*}_{sq}$, and $\epsilon^{*}_{ir}$ are the polarizations of the pump, signal, and idler respectively, summed over the indices o, q, and r.  If we have multiple crystals, of the same length, being pumped by the same beam, we can write the spatial dependence of the interaction Hamiltonian as the ``regular'' interaction Hamiltonian as convolution with a sum of delta functions.  We point out that this results merely in the sum of the individual interaction Hamiltonians, and we are considering a simplified case with some interesting results.  


\section{Maximization of the Biphoton Spectrum for a Determined Solid Angle}

In this section, we maximize the biphoton spectrum for a desired set of signal and idler k-vectors, and find the corresponding optimal expansion in the Elegant Gauss-Hermite Modes.  We will do this by invoking a variational principle and determining the structure of the pump spatial mode that will produce the largest probability of measuring a biphoton for an arbitrary set of signal and idler k-vectors.  As a result, we will also show that the only way to produce signal and idler k-vectors along the optic axis is to use the zeroth order Transverse Electromagnetic mode (we note that this is identical in the case of the CGH and regular GH modes).   

We start with our interaction Hamiltonian, assuming an initial vacuum mode for the signal and idler photon numbers:
\begin{align}
H_{SPDC} &=  -2h\sum_{n, m, o, q, r} c_{nm} \chi^{oqr} \epsilon_{po}\epsilon^{*}_{sq}\epsilon^{*}_{ir} V u_0  \int
d^3\nu_s d^3\nu_i\left\lbrace
\begin{aligned} 
&a^{\dagger}_{sq}(\nu_s) a^{\dagger}_{ir}(\nu_i) \sqrt{f_s(\nu_s) f_i(\nu_i)} (2\pi iw_0 \nu_{+x})^n\\
& \times (2\pi iw_0\nu_{+y})^m \Phi(\Delta \nu) e^{-\pi\lambda z_r\left(\nu_{+x}^2 + \nu_{+y}^2 \right)} 
\end{aligned} \right\rbrace
\end{align}

The integration will be performed over the detector volume, which we will assume is in the far field.  We thus wish to maximize the probability of the biphoton arriving at the detector.  We see that, assuming the interaction Hamiltonian is small and that the resulting measurement will be a two-photodiode correlation in the far field, which we may express as a finite integral over the k-vectors that the detectors are sensitive to, and approximate the resulting integral as slowly varying in the far field, and proportional to the functional form of the interaction Hamiltonian multiplied by the accompanying sensitive k-space volume of the detector (which will be labelled $\Delta(\nu_{s})^{3}$ and $\Delta(\nu_{i})^{3}$), the resulting term, state vector of interest will be proportional to:
\begin{align}
\ket{\psi} &\equiv H_{SPDC}\ket{0}\\
H_{SPDC} &=  -2h\sum_{n, m, o, q, r} c_{nm} \chi^{oqr} \epsilon_{po}\epsilon^{*}_{sq}\epsilon^{*}_{ir} V u_0  \int
d^3\nu_s d^3\nu_i\left\lbrace
\begin{aligned} 
&a^{\dagger}_{sq}(\nu_s) a^{\dagger}_{ir}(\nu_i) \sqrt{f_s(\nu_s) f_i(\nu_i)} (2\pi iw_0 \nu_{+x})^n\\
& \times (2\pi iw_0\nu_{+y})^m \Phi(\Delta \nu) e^{-\pi\lambda z_r\left(\nu_{+x}^2 + \nu_{+y}^2 \right)} 
\end{aligned} \right\rbrace\\
H_{SPDC} &\approx 2h\sum_{n, m, o, q, r} c_{nm} \chi^{oqr} \epsilon_{po}\epsilon^{*}_{sq}\epsilon^{*}_{ir} V u_0  
\Delta(\nu_s)^{3} \Delta(\nu_i)^{3}\left\lbrace
\begin{aligned} 
&a^{\dagger}_{sq}(\nu_s) a^{\dagger}_{ir}(\nu_i) \sqrt{f_s(\nu_s) f_i(\nu_i)} (2\pi iw_0 \nu_{+x})^n\\
& \times (2\pi iw_0\nu_{+y})^m \Phi(\Delta \nu) e^{-\pi\lambda z_r\left(\nu_{+x}^2 + \nu_{+y}^2 \right)} 
\end{aligned} \right\rbrace
\end{align}
Where the $\nu_{s}$ and $\nu_{i}$ terms are now a single value.

Any measurement that is to maximize the biphoton probability must maximize the corresponding bilinear term in $H_{SPDC}$.  If we dub our measurement operator ``M'', we find that:
\begin{align}
M &\propto \int d^{3}V \braket{\psi|a^{\dagger}_{s}a^{\dagger}_{i}a_{s}a_{i}|\psi}\\
M &\propto \sum_{nm}\sum_{ij} c^{*}_{nm} c_{ij}(-i w_{0}k_{+x})^{m}(-i w_{0}k_{+y})^{n}(i w_{0}k_{+x})^{i}(i w_{0}k_{+y})^{j}))
\end{align}

We thus wish to maximize the last term.  As we are working in the paraxial approximation, we may assume that $w_{0}k_{+x} <<<1$, and a similar assumption may be made in the y direction.  The maximization problem we must solve is also subject to one constraint, namely that the sum of all the pump spatial mode coefficients must be 1.  Explicitly:
\begin{align}
|c_{00}|^{2} + \sum_{n=1, m=1} |c_{nm}|^{2} &= 1\\
c_{00}c_{00}^{*} &= 1 - \sum_{nm}|c_{nm}|^{2}\\
c_{00}^{*}  &= \frac{1 - \sum_{nm}|c_{nm}|^{2}}{c_{00}}\\
\frac{\partial c_{00}^{*}}{\partial c_{00}} &=  -\frac{1 - \sum_{nm}|c_{nm}|^{2}}{c_{00}^{2}} = -\frac{c_{00}^{*}}{c_{00}}\\
\frac{\partial c_{00}^{*}}{\partial c_{nm}} &= -\frac{c_{nm}^{*}}{c_{00}}\\
\end{align}

We now minimize M, treating the mode coefficents and their complex conjugates as independent variables; this is equivalent to treating the real and imaginary parts as independent, as we may rewrite the mode coefficients as linear superpositions of the real and imaginary parts.  Explicityly, we institute: $\frac{\partial c_{ij}}{\partial c_{ij}^{*}} \equiv 0$.

We now proceed to find the critical points of M, noting that the sums over n,m, i, and j run from 1 to the maximum:
\begin{align}
M &= \sum_{mnij} \left(c_{00}^{*} + c_{nm}^{*} (-i w_{0}k_{+x})^{m}(i w_{0}k_{+y})^{n})\right)\left(c_{00} + c_{ij}(i w_{0}k_{+x})^{i}(i w_{0}k_{+y})^{j})\right)\\
\frac{\partial M}{\partial c_{00}} &= 0 =\sum_{mnij} c_{nm}^{*}\left((-i w_{0}k_{+x})^{m}(i w_{0}k_{+y})^{n})\right) - \frac{c_{00}^{*}}{c_{00}} c_{ij}\left((i w_{0}k_{+x})^{i}(i w_{0}k_{+y})^{j})\right)
\end{align}

We may set the phase of $c_{00}$ to zero without loss of generality.  As a consequence, we see that the term $c_{nm}^{*}\left((-i w_{0}k_{+x})^{m}(i w_{0}k_{+y})^{n})\right)$ must be real; this implies that we must set the phase of the modes to be  $(n+m)(-\pi / 2)$; this phase makes all terms in the interaction Hamiltonian real, and we may solve a simpler minimization problem; instituting the equaltiy $c_{nm} = (-i)^{n+m} \tilde{c}_{nm}$, we see that our problem now reduces to maximizing a real function squared, which is equivalent to maximizing the function:
\begin{align} 
M &= \left[\sum_{mn}\tilde{c}_{nm}\left((w_{0}k_{+x})^{m})(w_{0}k_{+y}^{n})\right) + c_{00}\right]^{2}
\end{align}

We now merely have to solve for the case of optimizing the term in the brackets to find:
\begin{align}
w_{0}k_{+x} &\equiv X\\
w_{0}k_{+y} &\equiv Y\\
c_{00} &= sqrt{1  - \sum_{nm}\tilde{c}_{nm}^{2}}\\
M' &= \sum_{mn}{\tilde{c}_{nm}X^{n}Y^{m} + c_{00}}\\
\frac{\partial M}{\partial \tilde{c}_{ij}} &= X^{i}Y^{j} - \tilde{c}_{ij} / c_{00}\\
\tilde{c}_{ij} &= X^{i}Y^{j} c_{00}\\
c_{00}^{2}(1 + \sum_{nm} X^{2m}Y^{2n}) &= 1\\
c_{00} &= \sqrt{\frac{1}{1 + \sum_{nm} X^{2m}Y^{2n}}}\\
\tilde{c}_{ij} &= \frac{X^{i}Y^{j}}{\sqrt{1 + \sum_{nm} X^{2m}Y^{2n}}}
\end{align}

With this solution, we now see that the optimal coefficients between the modes is:
\begin{align}
c_{ij} = \frac{(-iw_{0}k_{+x})^{i}(-iw_{0}k_{+y})^{j}}{\sqrt{1 + \sum_{nm} (w_{0}k_{+x})^{2m}(w_{0}k_{+y})^{2n}}}
\end{align}

We have thus maximized the biphoton spectra for a given set of k-vectors for signal and idler.  We would also point out that the result will hold irregardless of the limits of the integral in k-vector space, as the minimiziation was performed over the mode coefficients.

\section{Conclusion}

We have, for the first time (to the author's knowledge), determined the effect of multiple pump spatial modes on the resulting spectra of biphotons that result from the second order nonlinear spontanous interaction ("downconversion'').  We achieved this result by expanding the incoming pump field (which we treated as classical, or equivalently assumed was in a coherent state) in a basis of Siegman's Complex Gauss Hermite modes, which, as we have shown, have a defined Fourier Transform.  This results in, besides the normal gaussian phase factor that accompanies the Fresnel approximation in optics, a polynomial expansion in the sum of the signal and idler frequencies.  The form of the resulting biphoton spectra for type-II downconversion (e-oe or o-oe) is shown in \ref{Answer}.

\bibliography{EGHSPDC_1}

\end{document}